\documentclass[aip,rsi,reprint,graphicx]{revtex4-1} 

\usepackage {CJK}
 
\usepackage{amssymb}           
\usepackage{graphics,graphicx,amssymb,amsmath,amsfonts,verbatim}
\usepackage{epsfig}
\usepackage{subfigure}
\usepackage[subnum]{cases}
\usepackage{color}
\usepackage{lineno}

\usepackage{epstopdf}
\draft 

\begin{document}


\title{Evaluation of noise limits of a precision ADC for direct digital signal integration of magnetic measurements} 



\author{R.~Cavazzana}
\author{M.~Gottado}
\author{A.~Rigoni}
\affiliation{Consorzio RFX, Corso Stati Uniti 4, Padova 35127, Italy}


\date{\today}

\begin{abstract}
The characterization of laboratory plasma instabilities, magnetic reconnection and turbulence associated phenomena, require the simultaneous signal sampling from arrays of magnetic sensors (hundreds or even thousands) to obtain spatial resolution, with several hundred kHz for time resolution. Magnetic measurements based on pick-up (Mirnov) coil are quite common in experimental pulsed devices for plasma research, thanks to their simplicity and reliability.  Being the signal from this type of sensor proportional to the time variation of the magnetic field ($dB/dt$), it has to be time-integrated to recover the instant value of magnetic field. Depending on the required integration time usually either analog integrators or chopped integrators are used. However these solutions tend to limit the frequency bandwidth in the kHz range, they are not easy to design and build, and require additional fast channels to directly acquire the $dB/dt$ signal to recover the plasma dynamic features at higher frequencies. 
In this paper we evaluate the feasibility of using a direct single channel precision ADC to allow the simultaneous acquisition of $dB/dt$ measurements and to provide the integrated B measurement by means of digital integration on the new RFX-mod2 device. On this purpose we interfaced an existing ADC-module to a Xilinx Zynq FPGA, in order to evaluate the intrinsic noise and to investigate the feasible integration window of this configuration. The result opens the door to a compact, cost-effective and reliable acquisition system, usable for simultaneous real-time control and transient signal recording, scalable from tenths to thousands channels, applicable to a broad class of pulsed plasma experimental devices.
\end{abstract}


\maketitle 

\section{Introduction}\label{section:Intro}

RFX-mod~\cite{SONATO2003161}~\cite{2} is a medium size toroidal plasma multi-configuration machine (major radius  $R=2.0 m$, minor radius $a=0.46m$, operated up to 2 MA current reversed field pinch (RFP) configuration or 0.5 T tokamak). The experiment is now being further upgraded to RFX-mod2~\cite{Peruzzo2018}: one of the major step foreseen is a substantial improvement of the magnetic measurement system. The total number of the new magnetic pick-up coil sensors will be increased with the aim of improved spatial resolution~\cite{marchiori2017upgraded}. Moreover the sensors will be moved inside the vacuum vessel widening their usable signal bandwidth up to 200~kHz. These desirable characteristics imply demanding requirements for the acquisition system. 
The extension of the present analog integration system~\cite{pomaro2005transducers}, followed by two separate sets of ADC channels, one for precision off-line transient data sampling and one for real-time control, would be costly, requiring several rack cubicles and severely limiting the useful bandwidth. Thus a more compact and cost effective solution is being investigated. A possible one could be based on the ATCA-MIMO-ISOL~\cite{carvalho2010reconfigurable} architecture originally developed for the plasma column vertical stabilization of the JET tokamak experiment and already tested on RFX-mod~\cite{manduchi2012upgrade}. In principle this architecture could be suitable for performing the numeric integration need for real-time plasma control, while simultaneously recording the dB/dt signals suitable for the study of the fast MHD dynamic processes taking place into the plasma~\cite{zuin2009current}~\cite{innocente2014tearing}. The acquisition channels of this architecture are isolated input modules using precision 18-bit, 2~MSamples/s SAR ADC coupled to an FPGA which then routes the data flow to the processing/storage servers through a PCI-e interface. These characteristics of high sensitivity and high serial throughput, combined with the galvanic insulation of the input section of each module make this solution very attractive.
However for our purposes, the ATCA/PCI-e architecture is still expensive, complex to manage, and requires substantial software and firmware development. With this in mind we decided to switch to simpler and more flexible hosting interface, while trying to keep the existing isolated ADC module. The use of the Xilinx Zynq system on chip, integrating both FPGA and CPU cores, allows to set-up a self-contained acquisition board with a drastic simplification of the system, providing both the autonomous local storage for transient recording function, as well as the ability to send real-time decimated and integrated data to the control servers through standard gigabit Ethernet interface. 
To assess the feasibility of this proposal, we developed an interface adapter between the MIMO ISOL ADC module and a commercial Redpitaya Zynq board, and then we investigated the limits of the direct numeric integration technique with presently available electronic components.
The goal is to check to what extent a directly acquired signal from the RFX-mod2 probes can be digitally integrated. Even if applied to this specific case, the results obtained here could be used as guidelines for other similar applications as well. 
The paper is organized as follows: the dynamic range is defined at first from the assumption of the lowest and the highest signal expected. This has been set led by the experience of RFX-mod operation, using the collected data on the existing analog system as a baseline. The structure of the ADC and the FPGA interface developed to gather data are then briefly described. A characterization of the acquired signal is than provided to verify the feasibility of this approach.

\section{Expected signal range}\label{section:signal_range}

\begin{figure}[ht]
\centering
\includegraphics[width=0.49\textwidth]{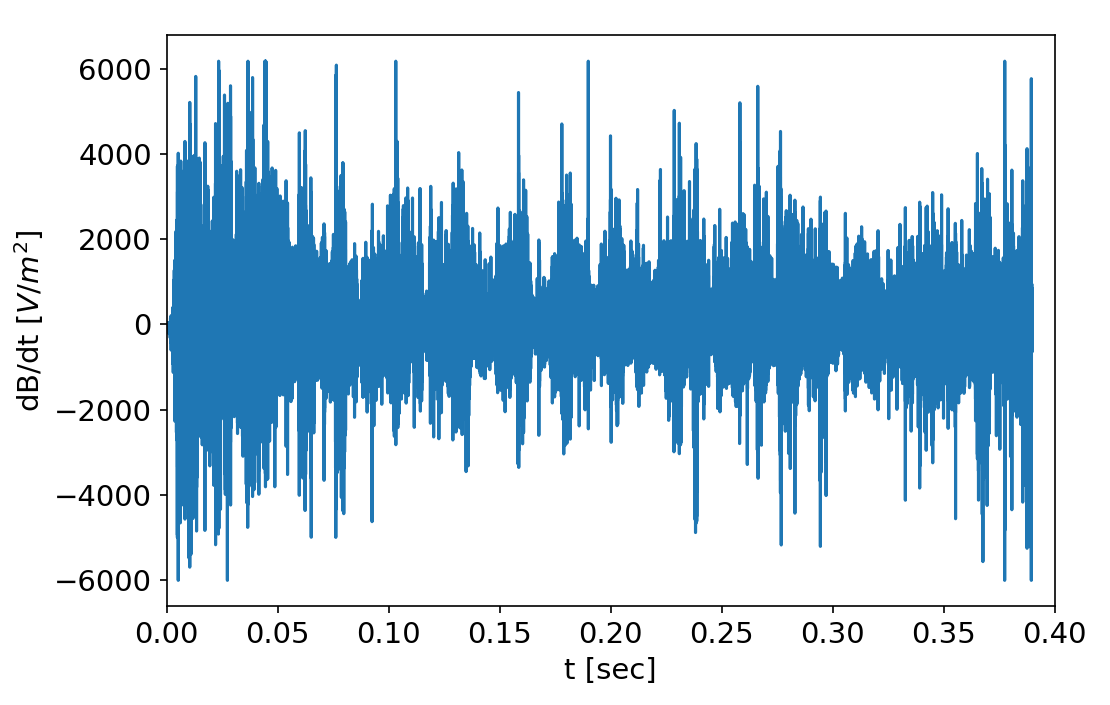}
\includegraphics[width=0.49\textwidth]{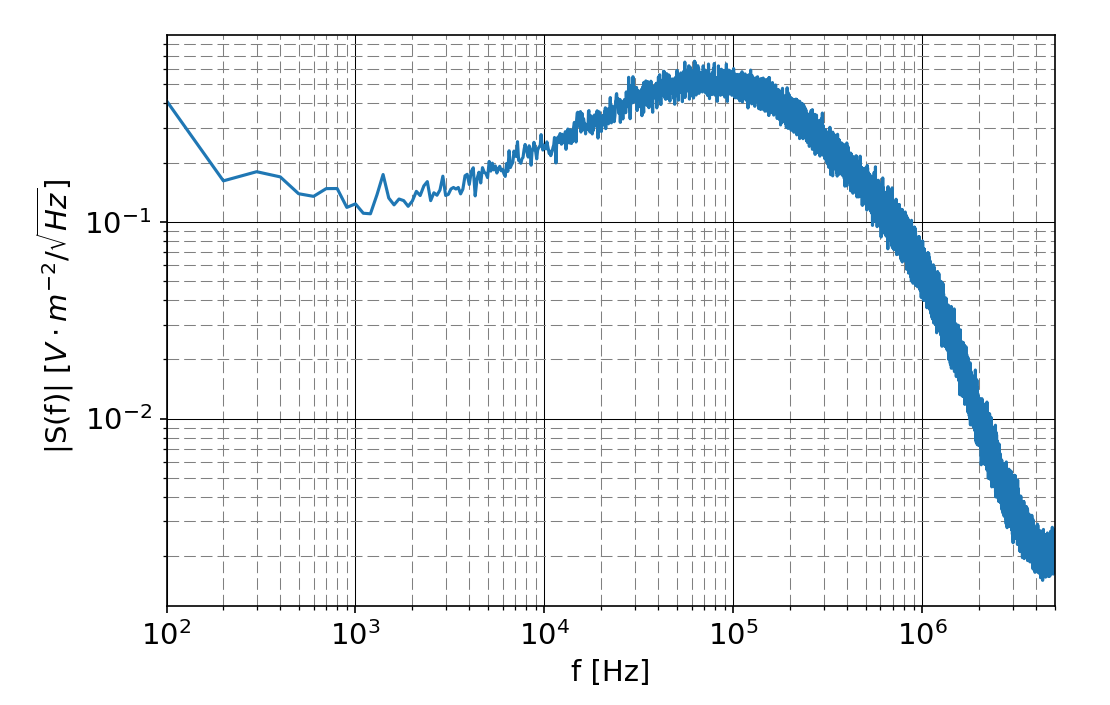}
\caption{dB/dt signal of the toroidal field component from an in-vessel wide bandwidth probe from a reversed field pinch pulse at 1.5 MA in RFXmod-2 (right signal, left spectrum).}
\label{fig:1}
\end{figure}

The dynamic range requirements for the signal acquisition system ask for a careful evaluation of both the minimum detectable and the maximum expected signal, keeping in mind that the signal has to be numerically integrated.  
The maximum level of the signals we are going to acquire can be straightforwardly derived from experimental measurements already carried out on RFX-mod, which can be used to estimate the level of the signal once the probe characteristics and its measurement chain are defined. 
The probes used up to now on RFX-mod~\cite{pomaro}~\cite{pomaro2005transducers} present a typical sensing area of 0,025 m2 and proved to offer a good compromise between compactness, signal sensitivity and usable bandwidth. The new probes of RFX-mod 2 will maintain these basic geometric and electric features, while the materials and the manufacturing process will be modified to comply with the vacuum installation~\cite{MARCHIORI2017892}. 
An example of the expected $dB/dt$ signal source and its spectrum, obtained by wide-band in-vessel probes installed on RFX-mod, are shown in Fig.\ref{fig:1}. This signal will be later used, to set the maximum signal level.

\begin{figure}[ht]
\centering
\includegraphics[width=0.49\textwidth]{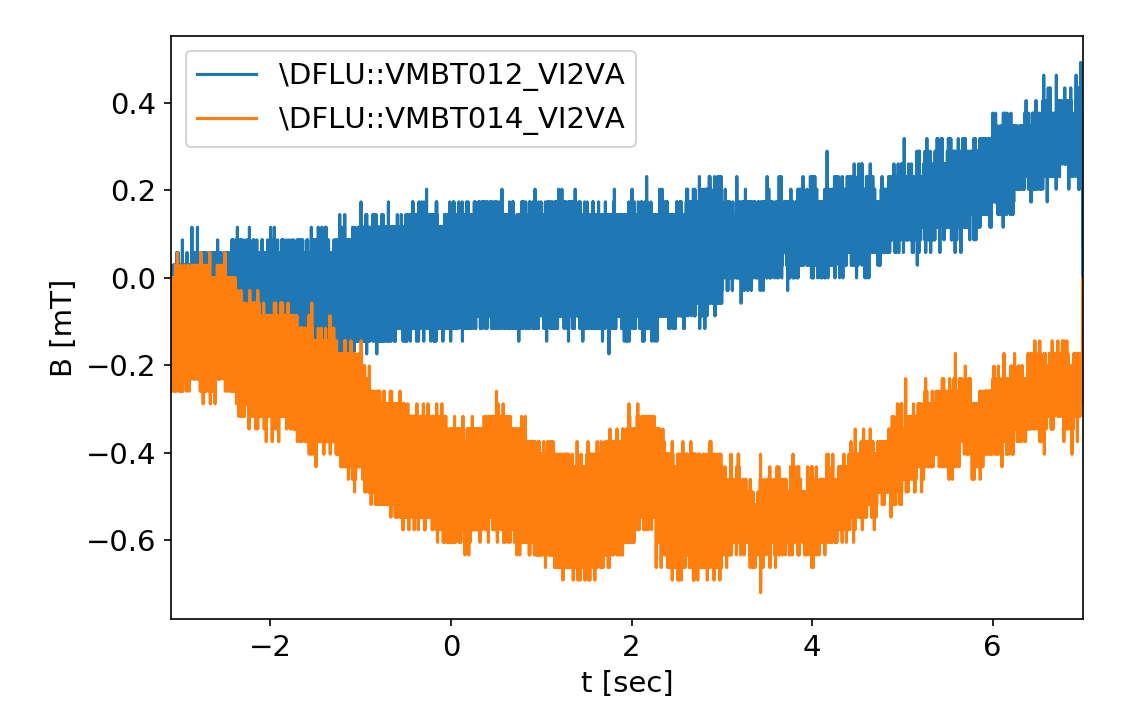}
\includegraphics[width=0.49\textwidth]{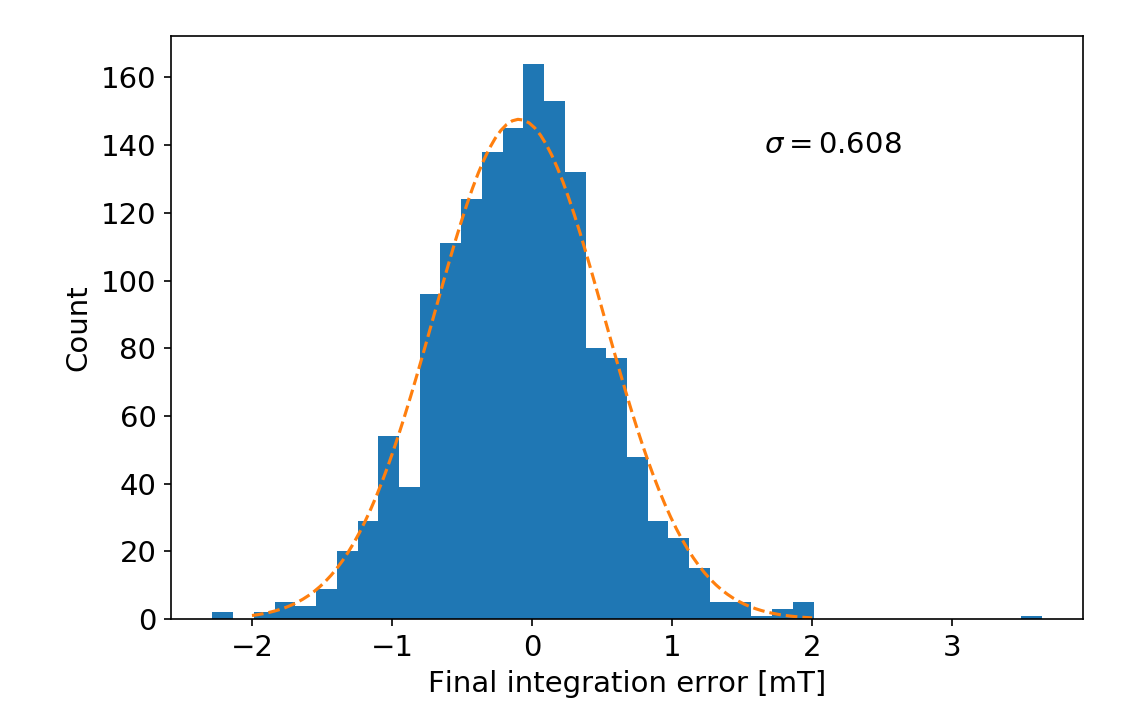}
\caption{Example of integrated signals (left) and distribution of final integration error of the analog integrators of RFX-mod after 10 sec, from toroidal field sensors on 8 “dry” pulses (1510 signals),(right).}
\label{fig:2}
\end{figure}

On the other hand the “floor” signal requirement involves the sensitivity to slowly changing magnetic field at low intensity. Magnetized plasmas are prone to amplify the spatial resonant component of an external field. Even an unwanted small field variation (the so called “field errors”) can have large effects on plasma behavior [compass scan]. The ability to detect these features is thus needed for example to allow the characterization of misalignments of the sensors and the structure of these small field components. 
As a comparison case, the performance of noise level and integrator drift of the actual conditioning system of RFX-mod is shown in Fig.~\ref{fig:2}. The final integration error is of the order of 0.6 mT, with the typical signal range to be measured which spans from 10 mT to 0.5 T. Most of the integration drift error can be corrected in post-processing, but it is necessary to keep it into an acceptable range to allow the use of these signals for real-time control purposes.

\begin{figure}[ht]
\centering
\includegraphics[width=0.49\textwidth]{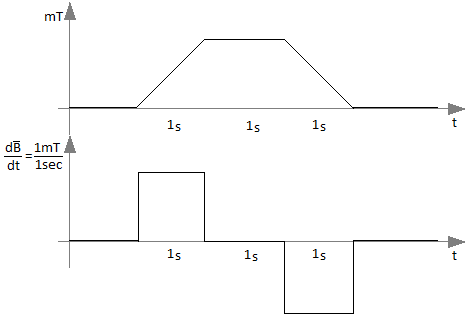}
\caption{desired minimum detectable signal scenario}
\label{fig:3}
\end{figure}

As the minimum detectable reference signal, we choose a trapezoidal 1 mT field with a slow trapezoidal evolution: 1 sec. ramp, 1 sec. state, 1 sec. ramp-down is shown on Fig. \ref{fig:3}.

The low signal level limit and the integration capability are severely affected by the $1/f$ noise of the circuit, which is poorly documented in the manufactures’ ADC data sheets and sometimes not well understood in the literature. These issues will be discussed later

\section{ADC module structure}\label{section:adc_module_structure}

\begin{figure}[ht]
\centering
\includegraphics[width=0.45\textwidth]{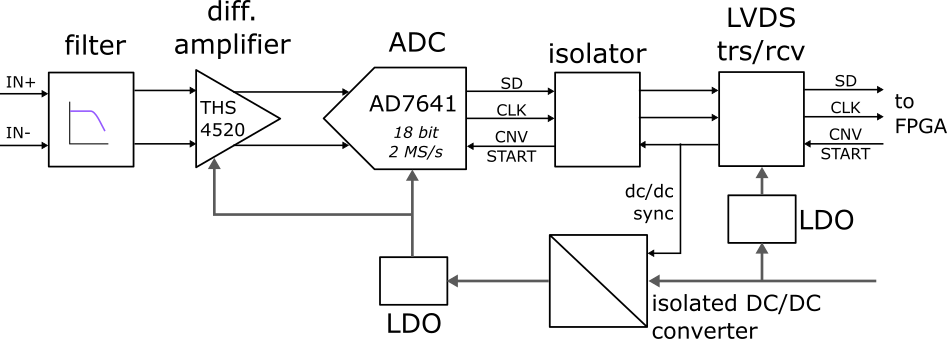}
\caption{block schematic of the ADC module}
\label{fig:4}
\end{figure}

The analog to digital conversion inside each of the ATCA-MIMO-ISOL board ADC modules is delegated to an Analog Devices AD7641, a 18-bit SAR converter that acquires signals from a fully differential input in the range 2.048 V at the maximum rate of 2~MSamples/s. A block scheme of the module is shown in Fig.~\ref{fig:4}: the analog input is initially filtered by a 1 pole, 100 kHz passive component connected to a differential amplifier THS4520 used as input range adapter; the AD7641 is configured to operate using serial communication protocol and the digital signals are delivered to the FPGA by means of a digital isolator. The power is supplied through an isolated DC/DC converter which is synchronized with the ADC conversion start command in order to minimize the noise spikes generated during power switching phase.

\section{ADC digital interface}\label{section:adc_digital_interface}

\begin{figure}[ht]
\centering
\includegraphics[width=0.45\textwidth]{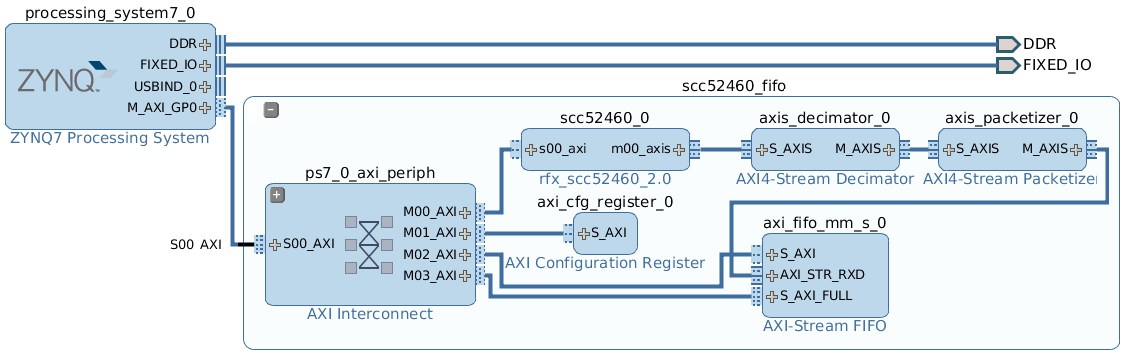}
\caption{block scheme of the implemented Zynq Xilinx AXI driver}
\label{fig:5}
\end{figure}

The module has been subsequently interfaced with a standard Xilinx Zynq FPGA board, developing a custom HDL code with the purpose of streaming data output to a proper embedded GNU Linux driver. Fig. \ref{fig:5} shows the schematic design of the bus connections of the FPGA logic.

Two main IP symbolic components are visible: the first one from the left of the sketch represents the processor entity, the second is related to the module digital acquisition. The processor unit presents two bus connections, externally linked to the DRR and fixed I/O hardware components, and one internally connected by the AXI\_GP0 general purpose bus to the other custom logics. Here all the IRQ and clock connections have been hidden to improve the schema readability but all the components are clocked by 125MHz Zynq internal multiplier and the FIFO drives an interrupt of the PS to inform the driver of the status changes. The custom logic named scc52460\_fifo encloses all the implemented design exploiting a 32-bit Xilinx AXI-Stream FIFO core with the input connected to a chain of custom cores performing in sequential order: the acquisition from module LVDS serial port, a configurable decimation of acquired data, and a configurable length packetization of the output stream. The communication protocol selected in the schematic hardware design of ADC module is the AD7641 serial master WARP mode connection with a further external flip-flop connected to the clock output that makes the signal transaction dual edge synchronized (rising and falling). This aims at achieving, likewise the DDR protocol specification, the same throughput with half of the frequency that in this way matches the one of the signal data.  The read implementation approach is then a double buffers with two process that have the sensibility to the respectively rising and falling edge of the SCLK\_in clock input port. On the other side a proper kernel module has been developed implementing the Xilinx FIFO core communication protocol~\cite{axi4_stream_fifo_2016}, in this way the kernel spools data from the FIFO buffer to the user allocated memory. Other solutions could be certainly applied to read data from the device, for example the Zynq is equipped with two high speed embedded DMA channels that are able to write data directly into DDR relieving the CPU of the mem-copy load. The proper transfer architecture should be always chosen in relation with the overall project complexity and goals. The system has been then tested for long lasting acquisition pulses at full speed 2 MSPS output and it resulted to be able to work properly in continuous mode.

\section{ADC module analog characteristics}\label{section:adc_module_analog_characteristics}

\begin{figure}[ht]
\centering
\includegraphics[width=0.49\textwidth]{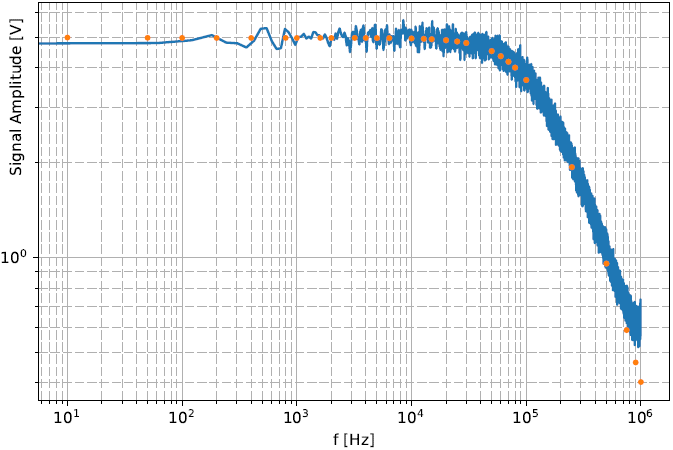}
\caption{Frequency response of ADC module (spectral levels are amplitude normalized).}
\label{fig:6}
\end{figure}

The module frequency response has been characterized by means of an HP33120A generator, with two different techniques: a) providing a series of sine waves at fixed frequency at 10 Vpp and b) with a white noise output. The result is shown in Fig. \ref{fig:6}, where the effect of the 100 kHz low pass input filter and the residual aliasing are clearly visible. In the following the Welch method has been used to compute an estimate of the power spectral density (WPSD i.e. Welch Power Spectral Density hereafter) by successively dividing the data into overlapping segments, computing a modified periodogram for each segment and finally averaging the obtained data. 

\begin{figure}[ht]
\centering
\includegraphics[width=0.49\textwidth]{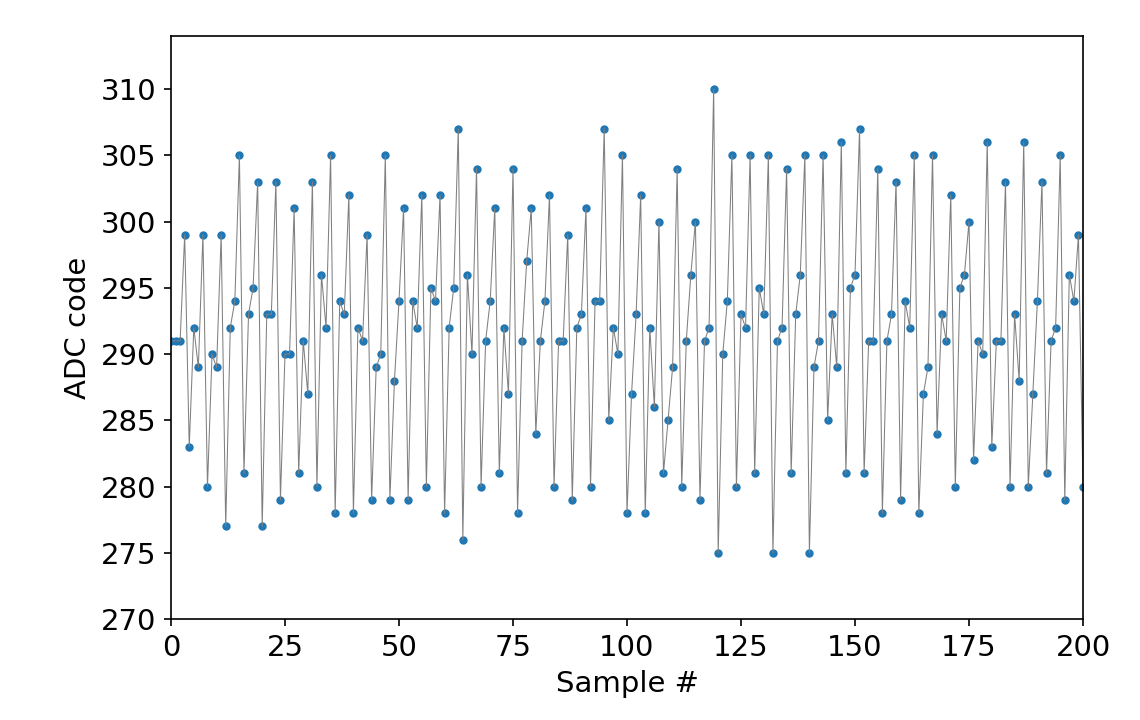}
\includegraphics[width=0.49\textwidth]{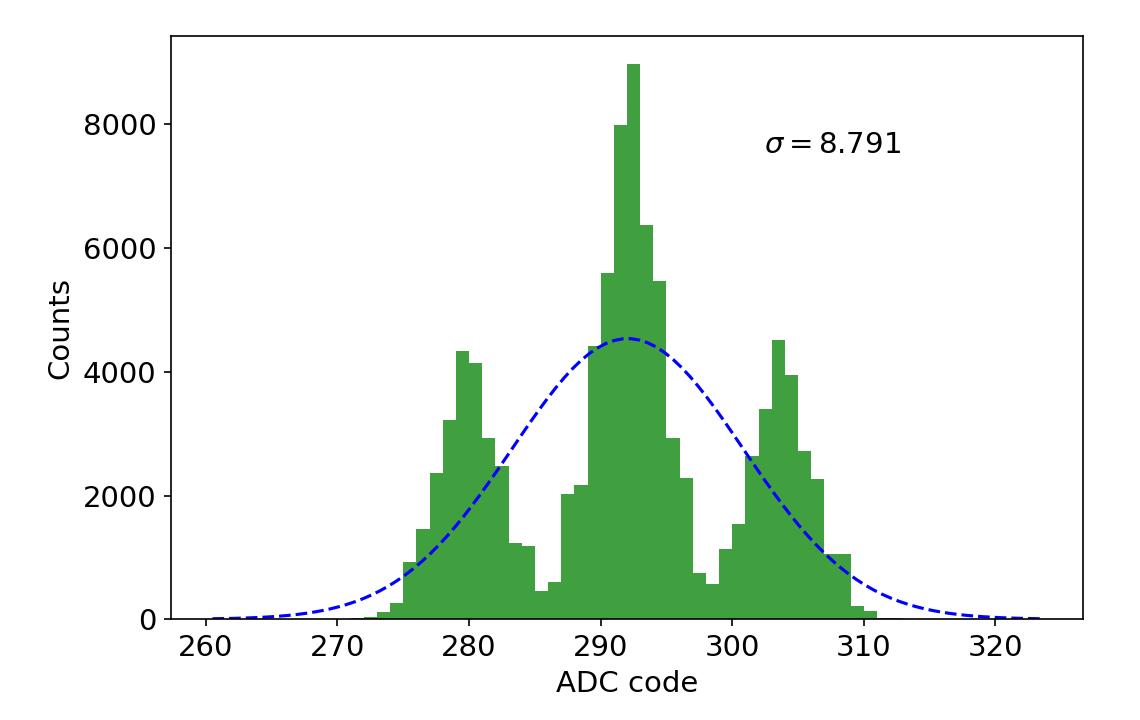}
\caption{Noise sampling (left) and noise histogram (100,000 samples, right) of the module in standard configuration with 50 Ohm terminated input. Note the repetitive synchronous spikes and the resulting multimodal distribution due to the DC/DC converter residual noise.}
\label{fig:7}
\end{figure}

\begin{figure}[ht]
\centering
\includegraphics[width=0.49\textwidth]{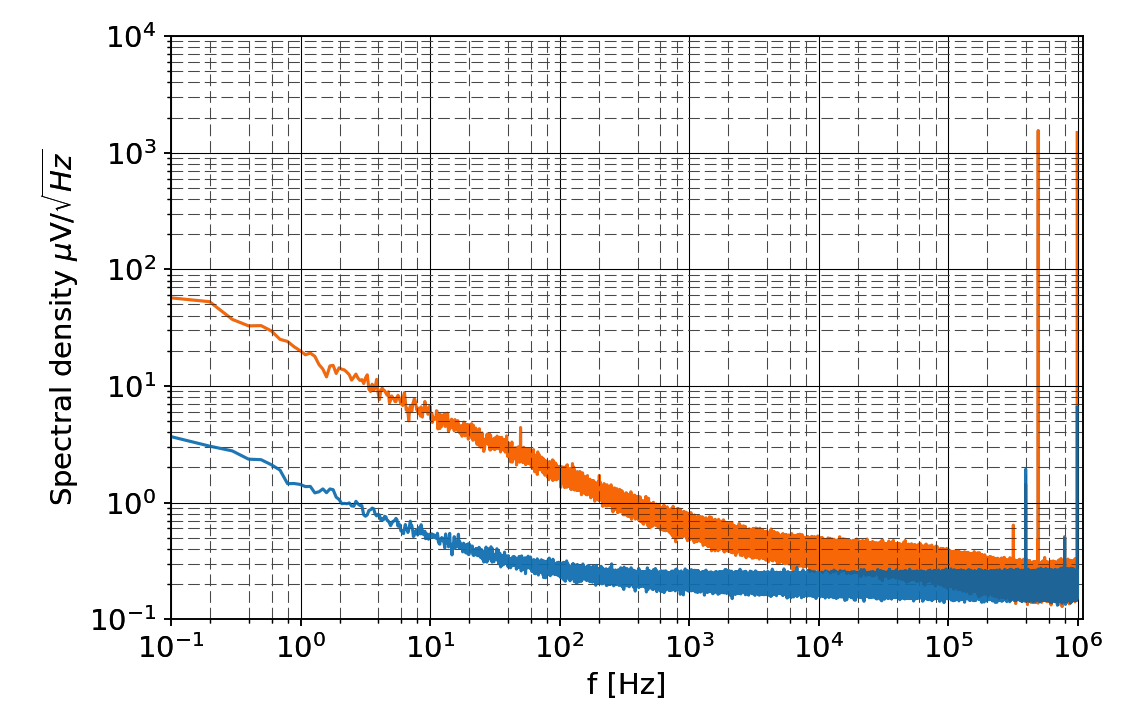}
\caption{Noise spectra (spectral levels are power normalized) after disconnecting the input amplifier; left panel circuit fed by its DC/DC converter, right panel by linear power supply – note the suppression of the frequency peak at 500 KHz.}
\label{fig:noise_spectra}
\end{figure}

In view of integrating the input signal, we carried out a detailed characterization of the noise seen by the ADC, for every one of the different configurations explored. Each run consisted of 8 noise snapshots, each one lasting 40 s (8∙107 samples). In order to reduce extraneous components, known noise sources have been either switched-off or kept far from the ADC board. The modules used are configured to have an input signal range of 10.3 V.
The noise has been characterized in time (Fig.~\ref{fig:7}) and frequency domains (Fig.~\ref{fig:noise_spectra}) marked with red line. The noise generated by the DC/DC converter is visible in the time domain as a series of synchronous spikes (\ref{fig:7}a), which are reflected in a histogram of the noise amplitude with three peaks (\ref{fig:7}b).

The frequency domain analysis shows two peaks due to isolation DC/DC converter, one at 500 kHz and its 1 MHz harmonic at Nyquist frequency, while at low frequency the spectrum is dominated by the 1/f noise. In order to better identify the noise sources inside the circuit, the DC/DC converter has been replaced by a linear power supply. Although the histogram of the sampled noise becomes comparable to the ADC specifications the spectrum improvement remains restricted to specific high frequencies that are easily filtered out.

\begin{figure}[ht]
\centering
\includegraphics[width=0.49\textwidth]{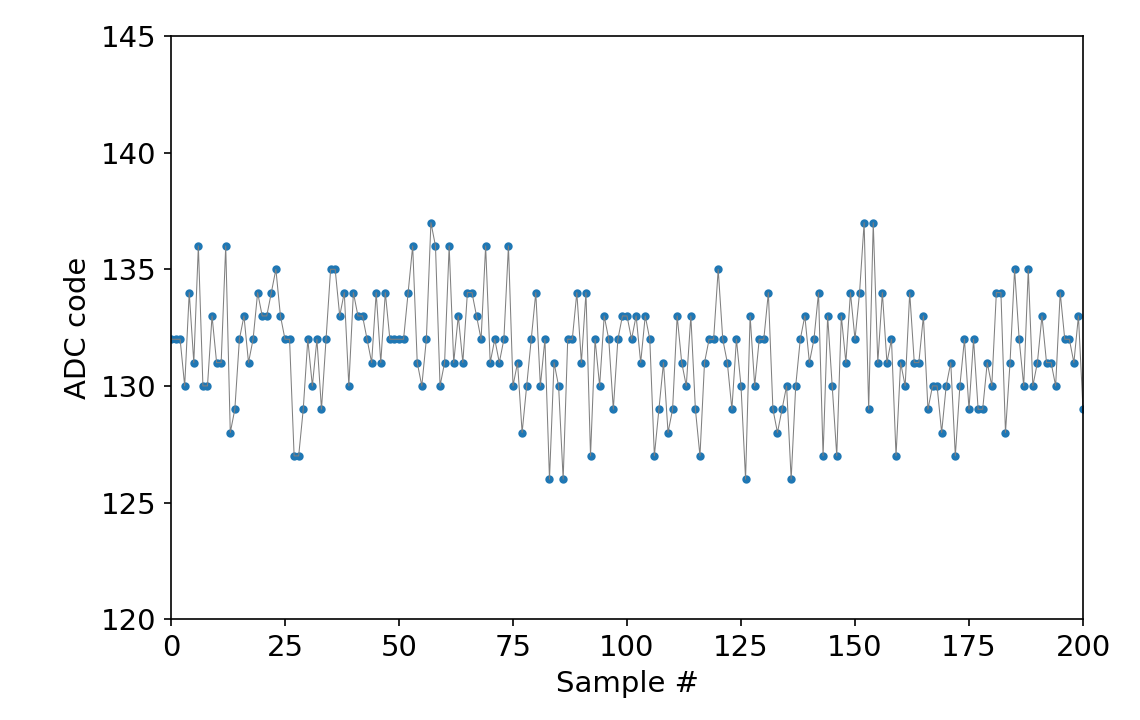}
\includegraphics[width=0.49\textwidth]{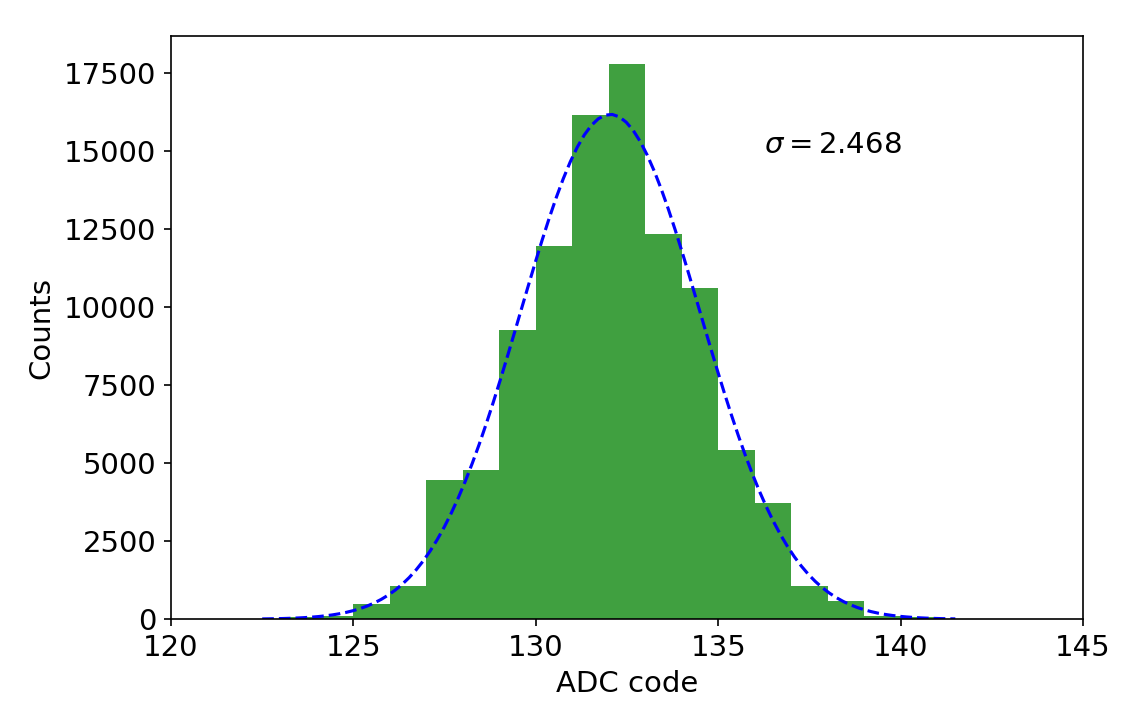}
\caption{Noise sampling (left) and noise histogram (100,000 samples, right) with 50 Ohm terminated input and the module fed by linear power supply.}
\label{fig:9}
\end{figure}


In the final circuit configuration the input amplifier has been removed as well, this time obtaining a reduction of the 1/f noise by a factor 20. A comparison of the overall noise spectrum with the final configuration is proposed in (Fig. \ref{fig:noise_spectra}), where the red spectrum represents the complete noise recorded without any intervention in the module, and the blue spectrum is the result of the signal acquisition with the DC/DC converter and the linear supply removed. 


\section{Integration of noise and expected drift error}\label{section:integration_of_noise_and_expected_drift}

As already mentioned in the chapter 1, one of the main goals of the present work is the complete replacement of existing analog integrators (Fig. \ref{fig:2}) with direct numerical integration of the digitally acquired signal. For short time integration, the noise spectrum involved is in the high frequency range, i.e. dominated by the white noise. In this case the numerical integration of a sampled sequence $s_i$ is a well defined case, being equivalent to a discrete random walk in which the distribution of the steps $s_i$ has zero mean with a finite variance $\sigma^2$. Thus given the final integral sum after $n$ steps $S_n =\sum_{i=0}^n s_i $, its variance is simply proportional to number of steps: $E|S_n^2| = \sigma^2 n $.

\begin{figure}[ht]
\centering
\includegraphics[width=0.49\textwidth]{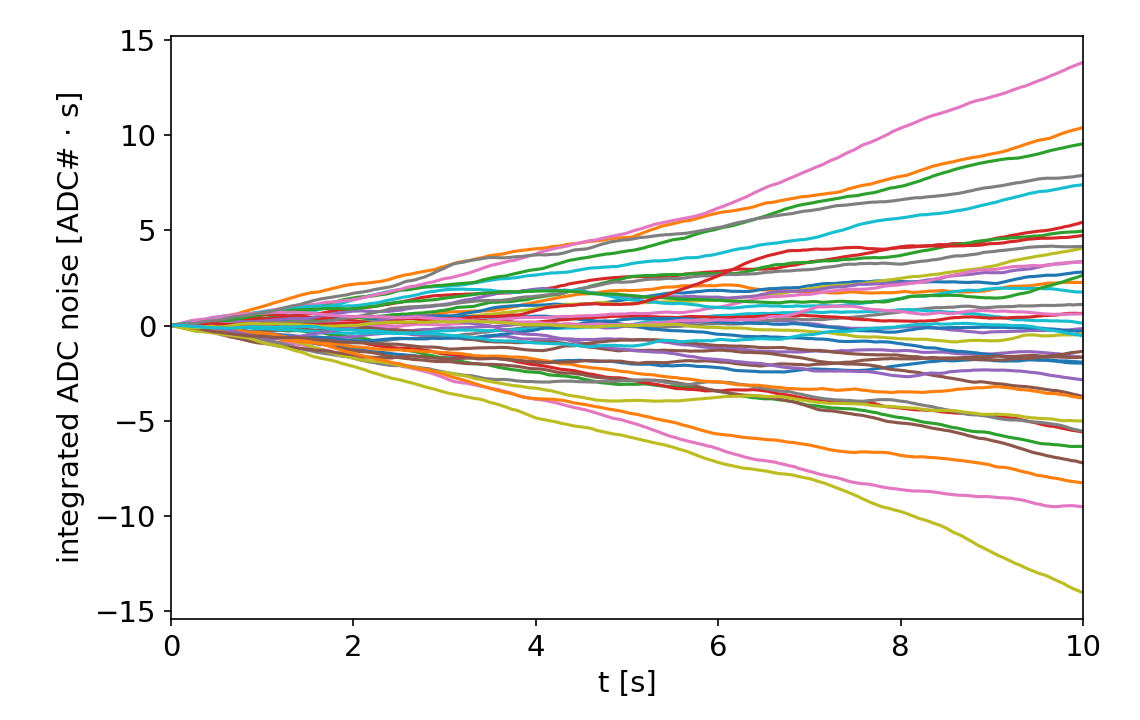}
\includegraphics[width=0.49\textwidth]{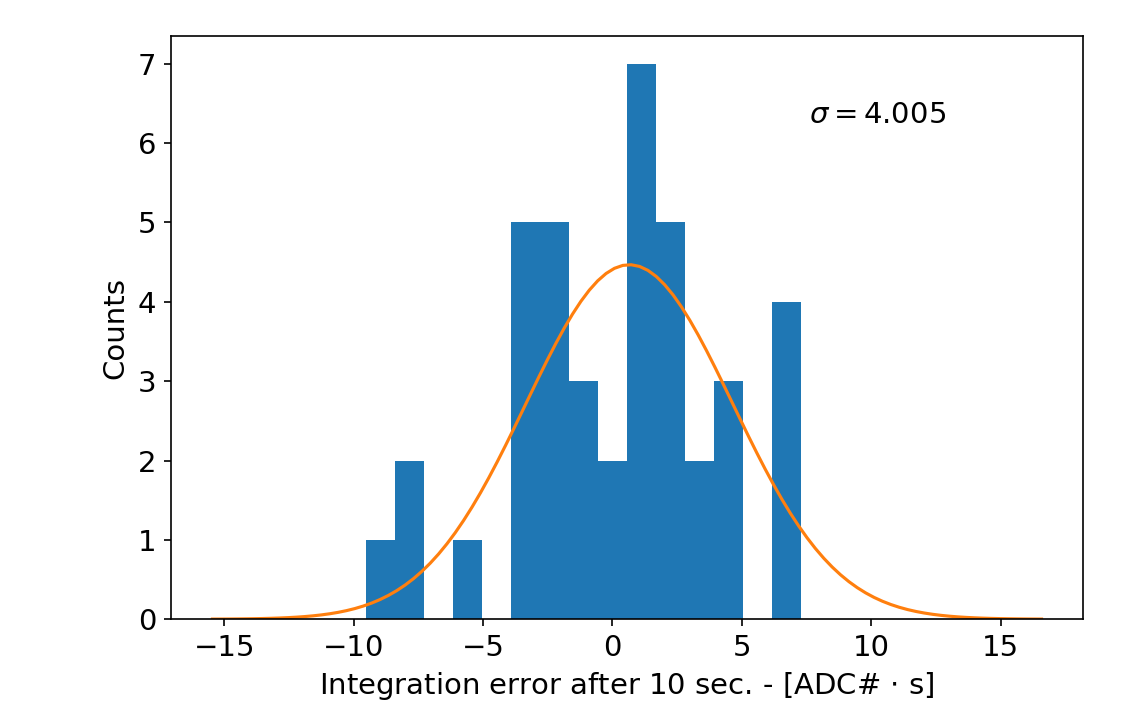}
\caption{Noise integrated signals from the module in standard configuration}
\label{fig:12}
\end{figure}

In the case at issue the integration time lasts several seconds and the desired frequency range extends far below the typical 1/f corner of standard electronic components. The integration is then dominated by the $1/f$ noise level. Contrary to the white noise, for this case there is no simple relation linking the noise level to the standard deviation of the final integration step. For this reason the problem has been tackled by means of a statistical analysis based on experimental data.
%
%
We performed the numerical integration on a number of samples [quanti?] on the different circuit configuration considered. As shown if Fig.~\ref{fig:12}, the estimated variance of the final value of the integrated signal after 10s with the ADC module in standard configuration, is about 4 ADC $\text{count}\cdot\text{sec}$. 

\begin{figure}[ht]
\centering
\includegraphics[width=0.49\textwidth]{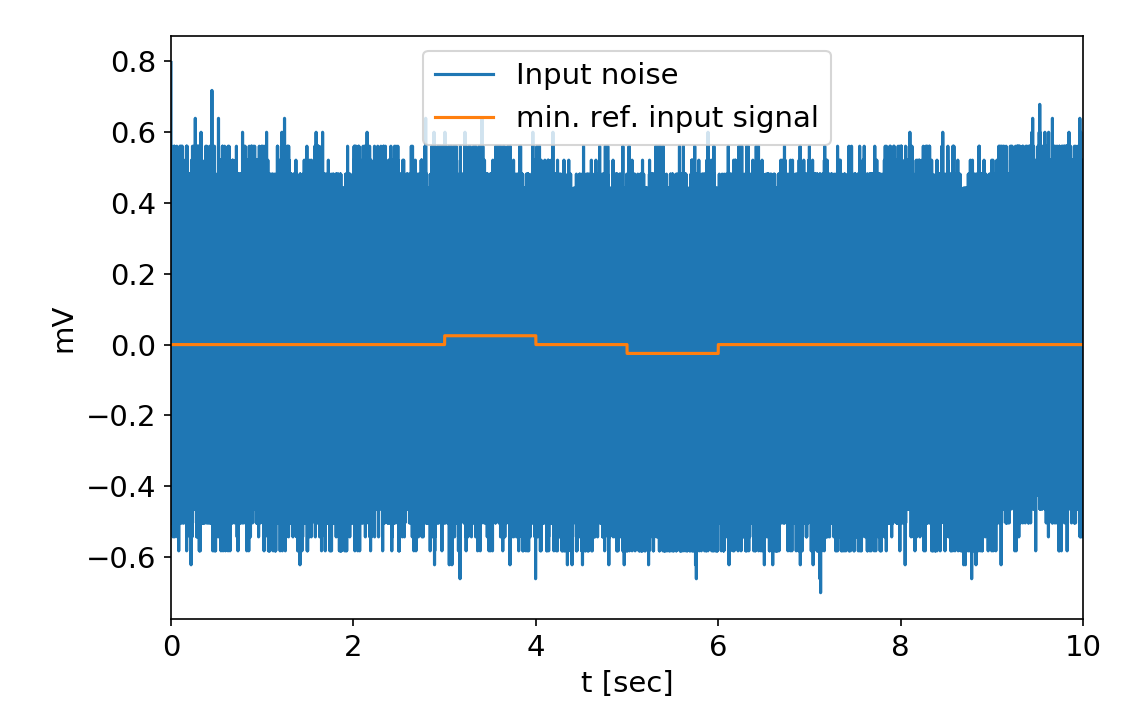}
\includegraphics[width=0.49\textwidth]{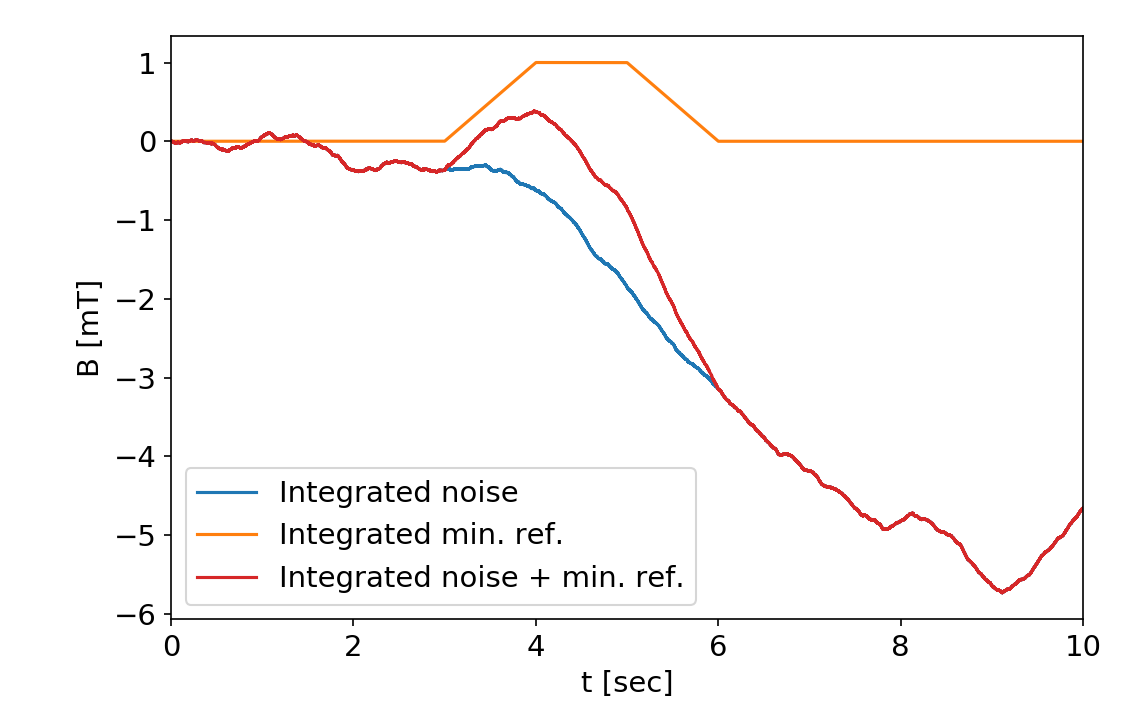}
\caption{Expected signals with a probe connected directly connected to the input module, with the minimum reference input signal. The signal appears indistinguishable from the noise.}
\label{fig:13}
\end{figure}

This translates in an equivalent final error for the magnetic probes of 5~mT. As appears from Fig.~\ref{fig:13}, besides its expected final value, the integration error applied to the reference magnetic measurement case makes practically impossible the reconstruction of the magnetic signal.

\begin{figure}[ht]
\centering
\includegraphics[width=0.49\textwidth]{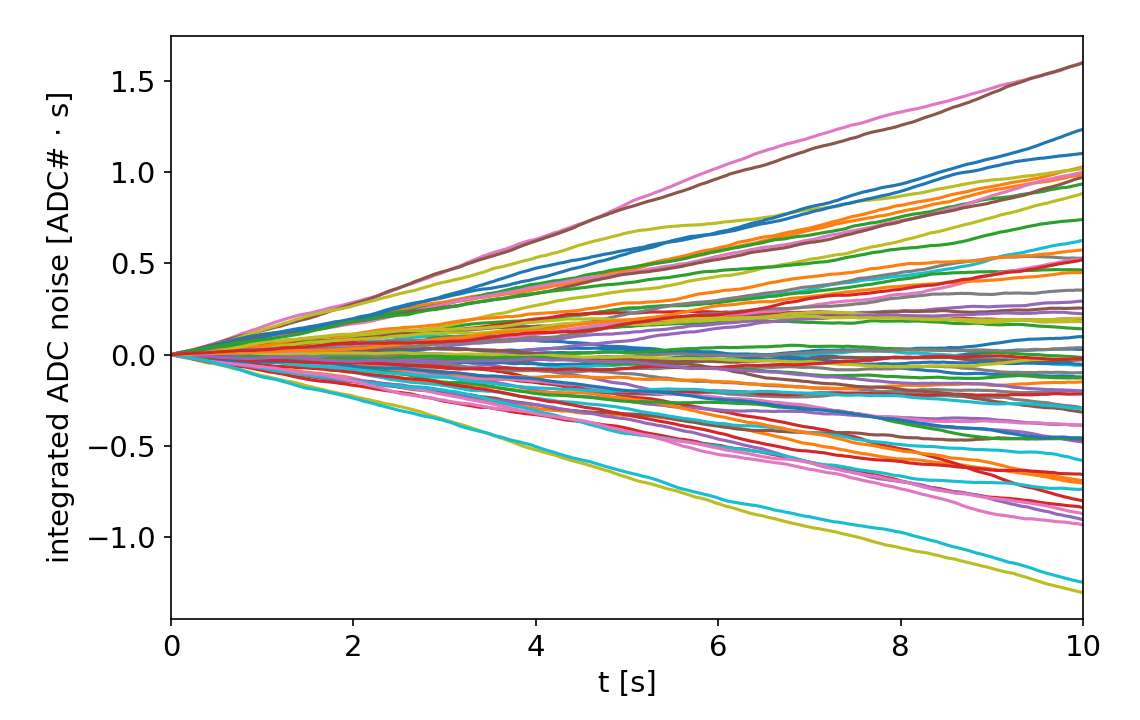}
\includegraphics[width=0.49\textwidth]{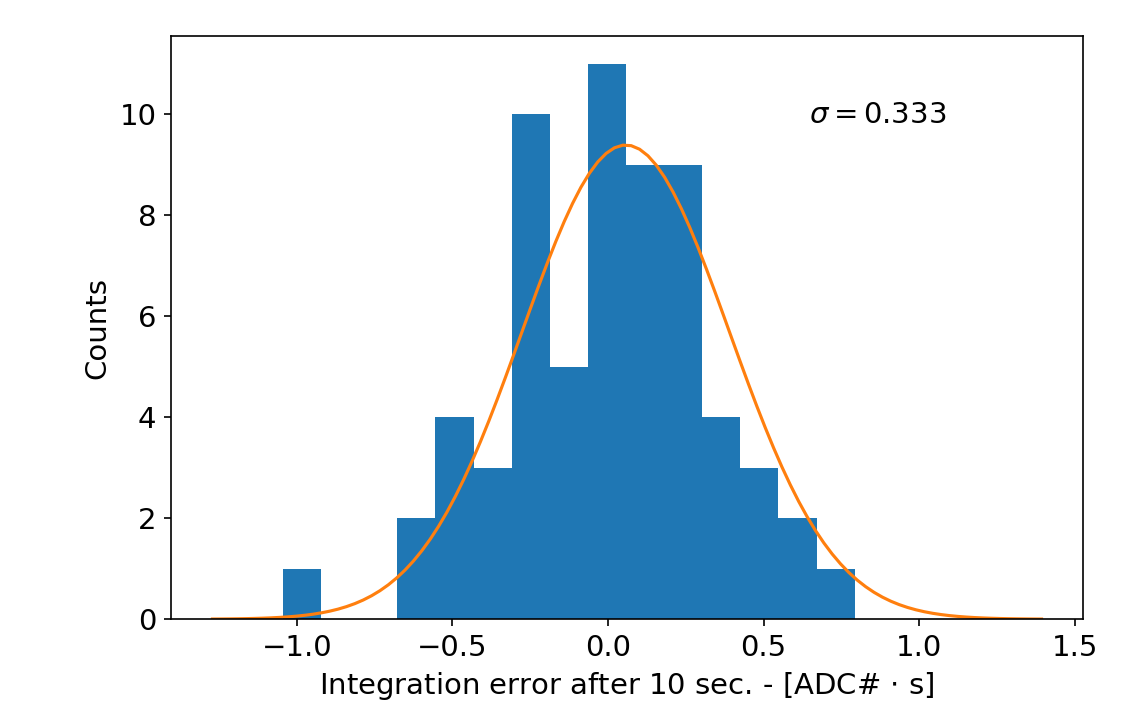}
\caption{Noise integrated signals from the module with the input amplifier disconnected, fed by the DC/DC converter.}
\label{fig:14}
\end{figure}

The case with the ADC without the input amplifier is reported in Fig. \ref{fig:14}, where the final error standard deviation is 0.33 ADC counts ∙ sec, which ten times lower than the previous case. 

\begin{figure}[ht]
\centering
\includegraphics[width=0.49\textwidth]{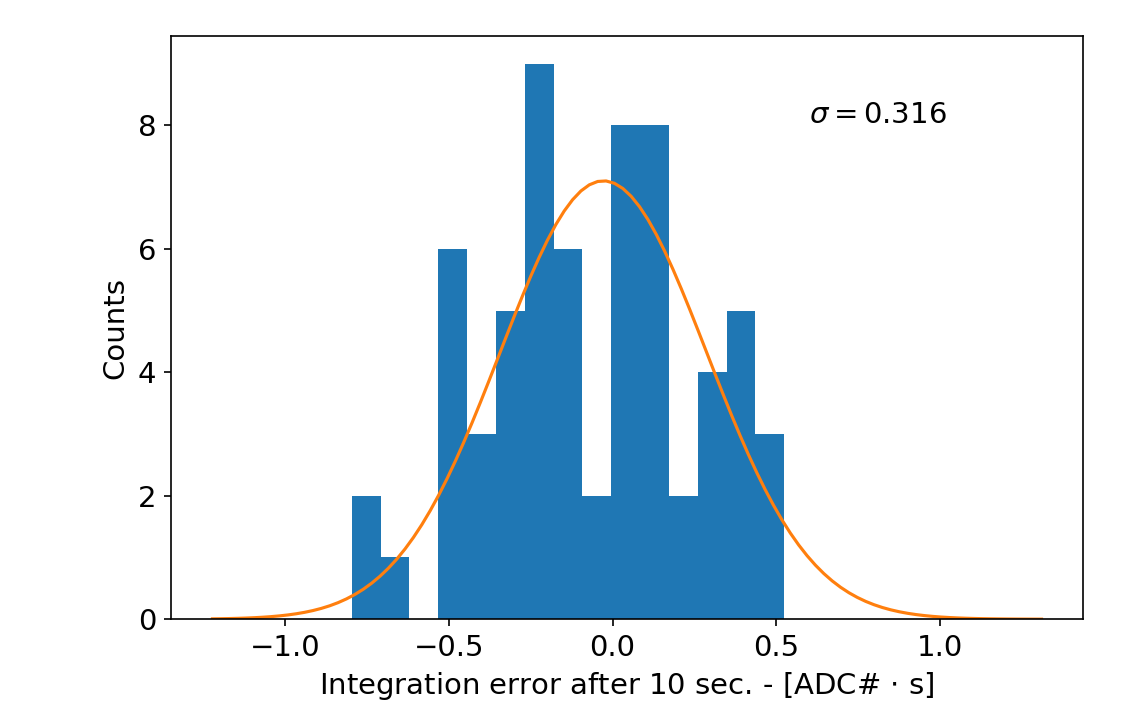}
\caption{Histogram of the noise integrated signals from the module with the input amplifier disconnected, fed by the linear power supply. Note that there is no practical gain in comparison with Fig. \ref{fig:14}.}
\label{fig:16}
\end{figure}

By adding the equivalent magnetic input reference signal to the noise, the desired signal can be successfully identified with an overall acceptable error and further corrected by suitable post processing.
~
\begin{figure}[ht]
\centering
\includegraphics[width=0.49\textwidth]{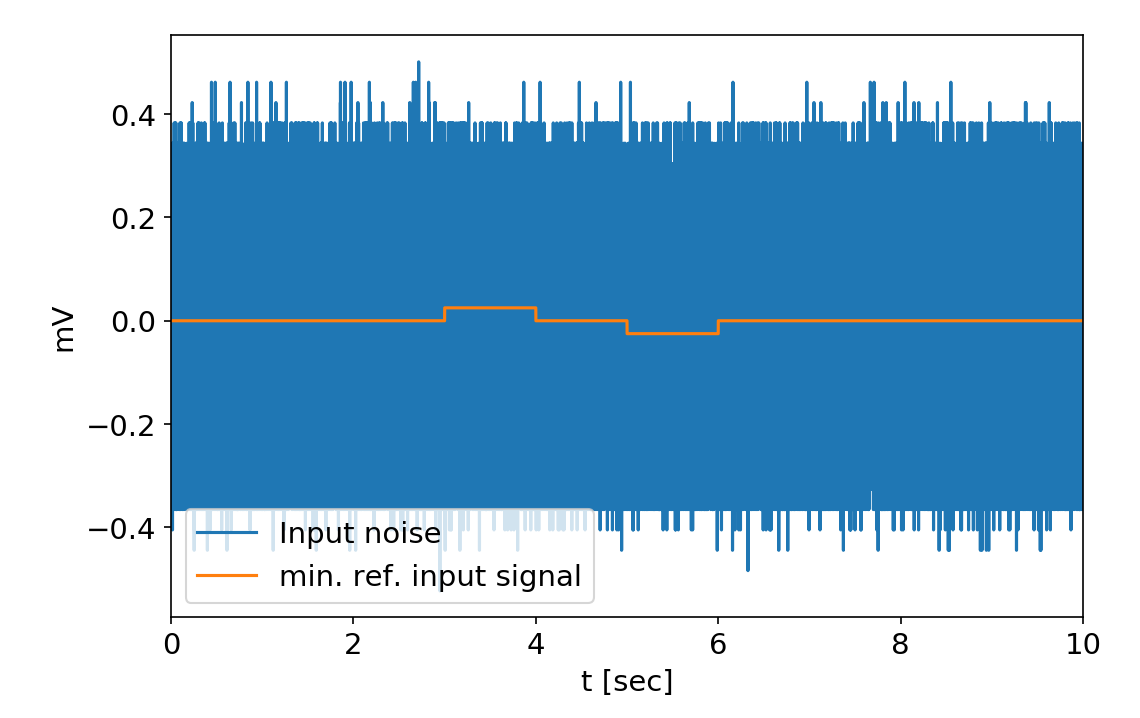}
\includegraphics[width=0.49\textwidth]{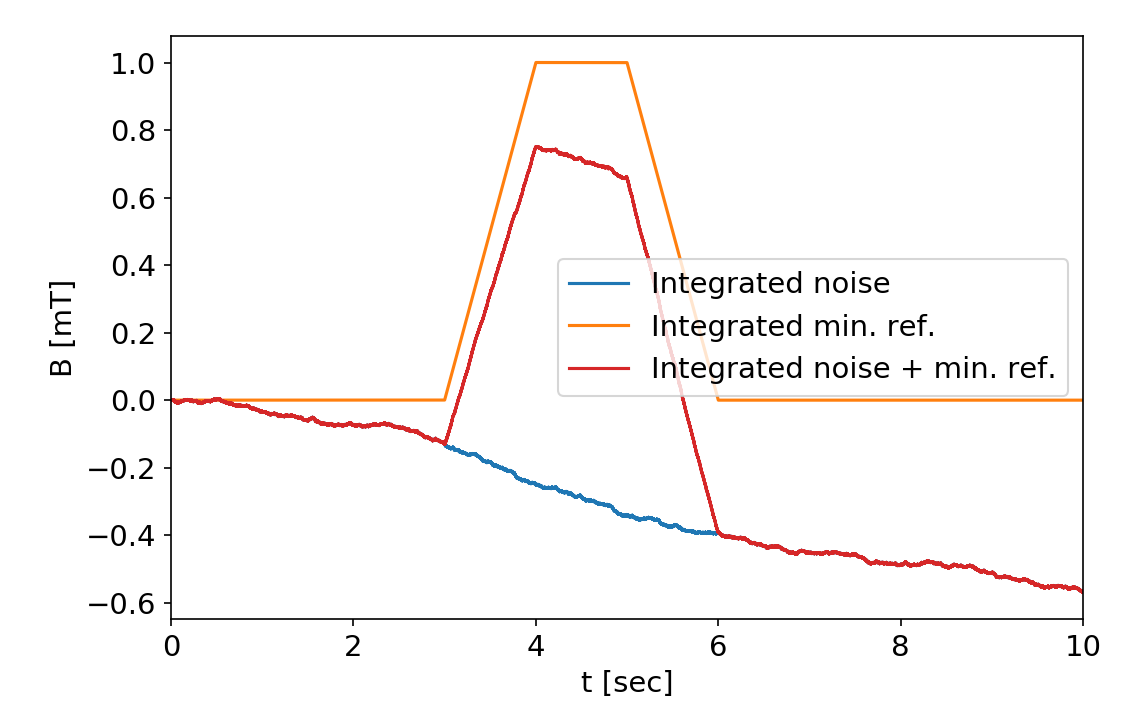}
\caption{Expected signals with the noise generated solely by the ADC input, with the minimum reference input signal. The integrated signal is now clearly visible.}
\label{fig:15}
\end{figure}
~
This last test reported, i.e. the ADC without any input amplifier and with a linear power supply, results in a standard deviation of 0.316 shown Fig.~\ref{fig:16}. This implies that for practical purposes the use of synchronous DC/DC converter does not affect the ADC conversion, at least with the purpose of numerical integration.

As briefly mentioned computational methods can be further applied to finally recover the actual input signal from a very low frequency drift. A commonly naive solution is to compute a standard linear interpolation of the noise evaluated outside the interval of interest, where the magnetic field is known to be steady, and subtract the interpolated offset during integration. Such operation can be easily applied on the signal in Fig. \ref{fig:15}, while is likely to fail on the case of Fig. \ref{fig:13}. This situation can be ascribed to the presence of 1/f noise components at relatively high frequencies, close to the signal band.


\section{Tests with magnetic probe}\label{section:tests_with_magnetic_probe}

As a final proof of concept the complete signal acquisition chain has been deployed in a test bench environment to check if circuit noise is compatible with the actual RFX probes. 
%
%
The test circuit used consists of a solenoid inducing a controlled field to the magnetic probe.
This was made by wrapping 40 turns of copper conductor of 2mm diameter in a coil that was estimated to present a 0.1mH inductance. In order to obtain the correct impedance to the circuit, a further resistive load of 2.5 Ohm was connected in series capable of dissipating a power of 100W, supplying in test by 4A it will dissipate 40W. The maximum current applicable to the total resistive load is 6.32A.


\begin{figure}[ht]
\centering
\includegraphics[width=0.49\textwidth]{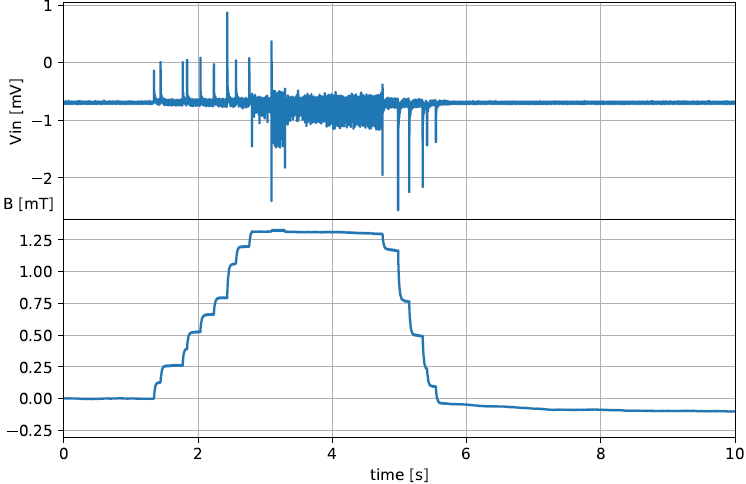}
\caption{manual }
\label{fig:18}
\end{figure}

The test result is shown in Fig \ref{fig:18}, where the signal provided to the magnetic probe was manually regulated using an external potentiometer changing the current flowing into the generator coil.
The direct acquired signal is reported on the curve at the high part of the image while the integrated result lies on the plot below. Here a successful magnetic field reconstruction can be directly appreciated measuring a 1.25~mT fluctuation within a time interval of about 3~s. The possible noise interpolation could exploit the leverage of the signal drift of about 1~mT measured along the entire 10 s signal to a reconstruction improvement of few $micro T$.




%
%
\section{Conclusions}\label{section:conclusions}

A complete acquisition chain has been built testing the feasibility of the numerical integration approach to recover the RFX magnetic field from solenoids probes. A complete isolated ADC module has been interfaced to an embedded SoC DAQ board exploiting a Xilinx FPGA real-time numerical elaboration.
In conclusion it has been shown that a direct digital integration appears feasible, even if it seems to require the state of the art components to recover high dynamic range and wide frequency spectrum signals at the same time, with a single channel. Both a simulated setup and an experimental testbed has been set up to evaluate the ADC noise impact on the integration results. 
Even using the best available commercial components, an accurate evaluation of the ADC characteristics must lead the choice of technology matching the desired input dynamic range and the bandwidth attenuation.  The need of high quality components has been shown to be related with the 1/f low drift noise mainly coming from the ADC and the input amplifier electronics. Although the 1/f noise seems to be the main issue for the integrated signals in long duration pulses, the high frequency oscillation due to the insulation DC/DC converter does not appear to create severe artifacts to the acquisition that can be reconstructed by digital filters synchronized with the acquisition system.

\clearpage 
\newpage


%
%

%



\bibliography{REC}

\begin{thebibliography}{12}%
\makeatletter
\providecommand \@ifxundefined [1]{%
 \@ifx{#1\undefined}
}%
\providecommand \@ifnum [1]{%
 \ifnum #1\expandafter \@firstoftwo
 \else \expandafter \@secondoftwo
 \fi
}%
\providecommand \@ifx [1]{%
 \ifx #1\expandafter \@firstoftwo
 \else \expandafter \@secondoftwo
 \fi
}%
\providecommand \natexlab [1]{#1}%
\providecommand \enquote  [1]{``#1''}%
\providecommand \bibnamefont  [1]{#1}%
\providecommand \bibfnamefont [1]{#1}%
\providecommand \citenamefont [1]{#1}%
\providecommand \href@noop [0]{\@secondoftwo}%
\providecommand \href [0]{\begingroup \@sanitize@url \@href}%
\providecommand \@href[1]{\@@startlink{#1}\@@href}%
\providecommand \@@href[1]{\endgroup#1\@@endlink}%
\providecommand \@sanitize@url [0]{\catcode `\\12\catcode `\$12\catcode
  `\&12\catcode `\#12\catcode `\^12\catcode `\_12\catcode `\%12\relax}%
\providecommand \@@startlink[1]{}%
\providecommand \@@endlink[0]{}%
\providecommand \url  [0]{\begingroup\@sanitize@url \@url }%
\providecommand \@url [1]{\endgroup\@href {#1}{\urlprefix }}%
\providecommand \urlprefix  [0]{URL }%
\providecommand \Eprint [0]{\href }%
\providecommand \doibase [0]{http://dx.doi.org/}%
\providecommand \selectlanguage [0]{\@gobble}%
\providecommand \bibinfo  [0]{\@secondoftwo}%
\providecommand \bibfield  [0]{\@secondoftwo}%
\providecommand \translation [1]{[#1]}%
\providecommand \BibitemOpen [0]{}%
\providecommand \bibitemStop [0]{}%
\providecommand \bibitemNoStop [0]{.\EOS\space}%
\providecommand \EOS [0]{\spacefactor3000\relax}%
\providecommand \BibitemShut  [1]{\csname bibitem#1\endcsname}%
\let\auto@bib@innerbib\@empty
\bibitem [{\citenamefont {Sonato}\ \emph {et~al.}(2003)\citenamefont {Sonato},
  \citenamefont {Chitarin}, \citenamefont {Zaccaria}, \citenamefont {Gnesotto},
  \citenamefont {Ortolani}, \citenamefont {Buffa}, \citenamefont {Bagatin},
  \citenamefont {Baker}, \citenamefont {Bello}, \citenamefont {Fiorentin},
  \citenamefont {Grando}, \citenamefont {Marchiori}, \citenamefont {Marcuzzi},
  \citenamefont {Masiello}, \citenamefont {Peruzzo}, \citenamefont {Pomaro},\
  and\ \citenamefont {Serianni}}]{SONATO2003161}%
  \BibitemOpen
  \bibfield  {author} {\bibinfo {author} {\bibfnamefont {P.}~\bibnamefont
  {Sonato}}, \bibinfo {author} {\bibfnamefont {G.}~\bibnamefont {Chitarin}},
  \bibinfo {author} {\bibfnamefont {P.}~\bibnamefont {Zaccaria}}, \bibinfo
  {author} {\bibfnamefont {F.}~\bibnamefont {Gnesotto}}, \bibinfo {author}
  {\bibfnamefont {S.}~\bibnamefont {Ortolani}}, \bibinfo {author}
  {\bibfnamefont {A.}~\bibnamefont {Buffa}}, \bibinfo {author} {\bibfnamefont
  {M.}~\bibnamefont {Bagatin}}, \bibinfo {author} {\bibfnamefont
  {W.}~\bibnamefont {Baker}}, \bibinfo {author} {\bibfnamefont {S.~D.}\
  \bibnamefont {Bello}}, \bibinfo {author} {\bibfnamefont {P.}~\bibnamefont
  {Fiorentin}}, \bibinfo {author} {\bibfnamefont {L.}~\bibnamefont {Grando}},
  \bibinfo {author} {\bibfnamefont {G.}~\bibnamefont {Marchiori}}, \bibinfo
  {author} {\bibfnamefont {D.}~\bibnamefont {Marcuzzi}}, \bibinfo {author}
  {\bibfnamefont {A.}~\bibnamefont {Masiello}}, \bibinfo {author}
  {\bibfnamefont {S.}~\bibnamefont {Peruzzo}}, \bibinfo {author} {\bibfnamefont
  {N.}~\bibnamefont {Pomaro}}, \ and\ \bibinfo {author} {\bibfnamefont
  {G.}~\bibnamefont {Serianni}},\ }\href {\doibase
  https://doi.org/10.1016/S0920-3796(03)00177-7} {\bibfield  {journal}
  {\bibinfo  {journal} {Fusion Engineering and Design}\ }\textbf {\bibinfo
  {volume} {66-68}},\ \bibinfo {pages} {161 } (\bibinfo {year} {2003})},\
  \bibinfo {note} {22nd Symposium on Fusion Technology}\BibitemShut {NoStop}%
\bibitem [{\citenamefont {Riva}\ \emph {et~al.}(2018)\citenamefont {Riva},
  \citenamefont {Vianello}, \citenamefont {Spolaore}, \citenamefont {Ricci},
  \citenamefont {Cavazzana}, \citenamefont {Marrelli},\ and\ \citenamefont
  {Spagnolo}}]{2}%
  \BibitemOpen
  \bibfield  {author} {\bibinfo {author} {\bibfnamefont {F.}~\bibnamefont
  {Riva}}, \bibinfo {author} {\bibfnamefont {N.}~\bibnamefont {Vianello}},
  \bibinfo {author} {\bibfnamefont {M.}~\bibnamefont {Spolaore}}, \bibinfo
  {author} {\bibfnamefont {P.}~\bibnamefont {Ricci}}, \bibinfo {author}
  {\bibfnamefont {R.}~\bibnamefont {Cavazzana}}, \bibinfo {author}
  {\bibfnamefont {L.}~\bibnamefont {Marrelli}}, \ and\ \bibinfo {author}
  {\bibfnamefont {S.}~\bibnamefont {Spagnolo}},\ }\href {\doibase
  10.1063/1.5008803} {\bibfield  {journal} {\bibinfo  {journal} {Physics of
  Plasmas}\ }\textbf {\bibinfo {volume} {25}},\ \bibinfo {pages} {022305}
  (\bibinfo {year} {2018})},\ \Eprint
  {http://arxiv.org/abs/https://doi.org/10.1063/1.5008803}
  {https://doi.org/10.1063/1.5008803} \BibitemShut {NoStop}%
\bibitem [{\citenamefont {Peruzzo}\ \emph {et~al.}(2018)\citenamefont
  {Peruzzo}, \citenamefont {Bernardi}, \citenamefont {Cavazzana}, \citenamefont
  {Bello}, \citenamefont {Palma}, \citenamefont {Grando}, \citenamefont
  {Perin}, \citenamefont {Piovan}, \citenamefont {Rizzolo}, \citenamefont
  {Rossetto}, \citenamefont {Ruaro}, \citenamefont {Siragusa}, \citenamefont
  {Sonato},\ and\ \citenamefont {Trevisan}}]{Peruzzo2018}%
  \BibitemOpen
  \bibfield  {author} {\bibinfo {author} {\bibfnamefont {S.}~\bibnamefont
  {Peruzzo}}, \bibinfo {author} {\bibfnamefont {M.}~\bibnamefont {Bernardi}},
  \bibinfo {author} {\bibfnamefont {R.}~\bibnamefont {Cavazzana}}, \bibinfo
  {author} {\bibfnamefont {S.~D.}\ \bibnamefont {Bello}}, \bibinfo {author}
  {\bibfnamefont {M.~D.}\ \bibnamefont {Palma}}, \bibinfo {author}
  {\bibfnamefont {L.}~\bibnamefont {Grando}}, \bibinfo {author} {\bibfnamefont
  {E.}~\bibnamefont {Perin}}, \bibinfo {author} {\bibfnamefont
  {R.}~\bibnamefont {Piovan}}, \bibinfo {author} {\bibfnamefont
  {A.}~\bibnamefont {Rizzolo}}, \bibinfo {author} {\bibfnamefont
  {F.}~\bibnamefont {Rossetto}}, \bibinfo {author} {\bibfnamefont
  {D.}~\bibnamefont {Ruaro}}, \bibinfo {author} {\bibfnamefont
  {M.}~\bibnamefont {Siragusa}}, \bibinfo {author} {\bibfnamefont
  {P.}~\bibnamefont {Sonato}}, \ and\ \bibinfo {author} {\bibfnamefont
  {L.}~\bibnamefont {Trevisan}},\ }\href {\doibase
  https://doi.org/10.1016/j.fusengdes.2018.05.066} {\bibfield  {journal}
  {\bibinfo  {journal} {Fusion Engineering and Design}\ ,\ } (\bibinfo {year}
  {2018})}\BibitemShut {NoStop}%
\bibitem [{\citenamefont {Marchiori}\ \emph
  {et~al.}(2017{\natexlab{a}})\citenamefont {Marchiori}, \citenamefont
  {Cavazzana}, \citenamefont {Bettini}, \citenamefont {Grando},\ and\
  \citenamefont {Peruzzo}}]{marchiori2017upgraded}%
  \BibitemOpen
  \bibfield  {author} {\bibinfo {author} {\bibfnamefont {G.}~\bibnamefont
  {Marchiori}}, \bibinfo {author} {\bibfnamefont {R.}~\bibnamefont
  {Cavazzana}}, \bibinfo {author} {\bibfnamefont {P.}~\bibnamefont {Bettini}},
  \bibinfo {author} {\bibfnamefont {L.}~\bibnamefont {Grando}}, \ and\ \bibinfo
  {author} {\bibfnamefont {S.}~\bibnamefont {Peruzzo}},\ }\href@noop {}
  {\bibfield  {journal} {\bibinfo  {journal} {Fusion Engineering and Design}\
  }\textbf {\bibinfo {volume} {123}},\ \bibinfo {pages} {892} (\bibinfo {year}
  {2017}{\natexlab{a}})}\BibitemShut {NoStop}%
\bibitem [{\citenamefont {Pomaro}\ and\ \citenamefont
  {Basso}(2005)}]{pomaro2005transducers}%
  \BibitemOpen
  \bibfield  {author} {\bibinfo {author} {\bibfnamefont {N.}~\bibnamefont
  {Pomaro}}\ and\ \bibinfo {author} {\bibfnamefont {F.}~\bibnamefont {Basso}},\
  }\href@noop {} {\bibfield  {journal} {\bibinfo  {journal} {Fusion engineering
  and design}\ }\textbf {\bibinfo {volume} {74}},\ \bibinfo {pages} {721}
  (\bibinfo {year} {2005})}\BibitemShut {NoStop}%
\bibitem [{\citenamefont {Carvalho}\ \emph {et~al.}(2010)\citenamefont
  {Carvalho}, \citenamefont {Batista}, \citenamefont {Correia}, \citenamefont
  {Neto}, \citenamefont {Fernandes}, \citenamefont {Gon{\c{c}}alves},\ and\
  \citenamefont {Sousa}}]{carvalho2010reconfigurable}%
  \BibitemOpen
  \bibfield  {author} {\bibinfo {author} {\bibfnamefont {B.}~\bibnamefont
  {Carvalho}}, \bibinfo {author} {\bibfnamefont {A.}~\bibnamefont {Batista}},
  \bibinfo {author} {\bibfnamefont {M.}~\bibnamefont {Correia}}, \bibinfo
  {author} {\bibfnamefont {A.}~\bibnamefont {Neto}}, \bibinfo {author}
  {\bibfnamefont {H.}~\bibnamefont {Fernandes}}, \bibinfo {author}
  {\bibfnamefont {B.}~\bibnamefont {Gon{\c{c}}alves}}, \ and\ \bibinfo {author}
  {\bibfnamefont {J.}~\bibnamefont {Sousa}},\ }\href@noop {} {\bibfield
  {journal} {\bibinfo  {journal} {Fusion Engineering and Design}\ }\textbf
  {\bibinfo {volume} {85}},\ \bibinfo {pages} {298} (\bibinfo {year}
  {2010})}\BibitemShut {NoStop}%
\bibitem [{\citenamefont {Manduchi}\ \emph {et~al.}(2012)\citenamefont
  {Manduchi}, \citenamefont {Barbalace}, \citenamefont {Luchetta},
  \citenamefont {Soppelsa}, \citenamefont {Taliercio},\ and\ \citenamefont
  {Zampiva}}]{manduchi2012upgrade}%
  \BibitemOpen
  \bibfield  {author} {\bibinfo {author} {\bibfnamefont {G.}~\bibnamefont
  {Manduchi}}, \bibinfo {author} {\bibfnamefont {A.}~\bibnamefont {Barbalace}},
  \bibinfo {author} {\bibfnamefont {A.}~\bibnamefont {Luchetta}}, \bibinfo
  {author} {\bibfnamefont {A.}~\bibnamefont {Soppelsa}}, \bibinfo {author}
  {\bibfnamefont {C.}~\bibnamefont {Taliercio}}, \ and\ \bibinfo {author}
  {\bibfnamefont {E.}~\bibnamefont {Zampiva}},\ }\href@noop {} {\bibfield
  {journal} {\bibinfo  {journal} {Fusion Engineering and Design}\ }\textbf
  {\bibinfo {volume} {87}},\ \bibinfo {pages} {1907} (\bibinfo {year}
  {2012})}\BibitemShut {NoStop}%
\bibitem [{\citenamefont {Zuin}\ \emph {et~al.}(2009)\citenamefont {Zuin},
  \citenamefont {Vianello}, \citenamefont {Spolaore}, \citenamefont {Antoni},
  \citenamefont {Bolzonella}, \citenamefont {Cavazzana}, \citenamefont
  {Martines}, \citenamefont {Serianni},\ and\ \citenamefont
  {Terranova}}]{zuin2009current}%
  \BibitemOpen
  \bibfield  {author} {\bibinfo {author} {\bibfnamefont {M.}~\bibnamefont
  {Zuin}}, \bibinfo {author} {\bibfnamefont {N.}~\bibnamefont {Vianello}},
  \bibinfo {author} {\bibfnamefont {M.}~\bibnamefont {Spolaore}}, \bibinfo
  {author} {\bibfnamefont {V.}~\bibnamefont {Antoni}}, \bibinfo {author}
  {\bibfnamefont {T.}~\bibnamefont {Bolzonella}}, \bibinfo {author}
  {\bibfnamefont {R.}~\bibnamefont {Cavazzana}}, \bibinfo {author}
  {\bibfnamefont {E.}~\bibnamefont {Martines}}, \bibinfo {author}
  {\bibfnamefont {G.}~\bibnamefont {Serianni}}, \ and\ \bibinfo {author}
  {\bibfnamefont {D.}~\bibnamefont {Terranova}},\ }\href@noop {} {\bibfield
  {journal} {\bibinfo  {journal} {Plasma Physics and Controlled Fusion}\
  }\textbf {\bibinfo {volume} {51}},\ \bibinfo {pages} {035012} (\bibinfo
  {year} {2009})}\BibitemShut {NoStop}%
\bibitem [{\citenamefont {Innocente}\ \emph {et~al.}(2014)\citenamefont
  {Innocente}, \citenamefont {Zanca}, \citenamefont {Zuin}, \citenamefont
  {Bolzonella},\ and\ \citenamefont {Zaniol}}]{innocente2014tearing}%
  \BibitemOpen
  \bibfield  {author} {\bibinfo {author} {\bibfnamefont {P.}~\bibnamefont
  {Innocente}}, \bibinfo {author} {\bibfnamefont {P.}~\bibnamefont {Zanca}},
  \bibinfo {author} {\bibfnamefont {M.}~\bibnamefont {Zuin}}, \bibinfo {author}
  {\bibfnamefont {T.}~\bibnamefont {Bolzonella}}, \ and\ \bibinfo {author}
  {\bibfnamefont {B.}~\bibnamefont {Zaniol}},\ }\href@noop {} {\bibfield
  {journal} {\bibinfo  {journal} {Nuclear Fusion}\ }\textbf {\bibinfo {volume}
  {54}},\ \bibinfo {pages} {122001} (\bibinfo {year} {2014})}\BibitemShut
  {NoStop}%
\bibitem [{\citenamefont {Bolzonella}\ \emph {et~al.}(2003)\citenamefont
  {Bolzonella}, \citenamefont {Pomaro}, \citenamefont {Serianni},\ and\
  \citenamefont {Marcuzzi}}]{pomaro}%
  \BibitemOpen
  \bibfield  {author} {\bibinfo {author} {\bibfnamefont {T.}~\bibnamefont
  {Bolzonella}}, \bibinfo {author} {\bibfnamefont {N.}~\bibnamefont {Pomaro}},
  \bibinfo {author} {\bibfnamefont {G.}~\bibnamefont {Serianni}}, \ and\
  \bibinfo {author} {\bibfnamefont {D.}~\bibnamefont {Marcuzzi}},\ }\href
  {\doibase 10.1063/1.1526930} {\bibfield  {journal} {\bibinfo  {journal}
  {Review of Scientific Instruments}\ }\textbf {\bibinfo {volume} {74}},\
  \bibinfo {pages} {1554} (\bibinfo {year} {2003})}\BibitemShut {NoStop}%
\bibitem [{\citenamefont {Marchiori}\ \emph
  {et~al.}(2017{\natexlab{b}})\citenamefont {Marchiori}, \citenamefont
  {Cavazzana}, \citenamefont {Bettini}, \citenamefont {Grando},\ and\
  \citenamefont {Peruzzo}}]{MARCHIORI2017892}%
  \BibitemOpen
  \bibfield  {author} {\bibinfo {author} {\bibfnamefont {G.}~\bibnamefont
  {Marchiori}}, \bibinfo {author} {\bibfnamefont {R.}~\bibnamefont
  {Cavazzana}}, \bibinfo {author} {\bibfnamefont {P.}~\bibnamefont {Bettini}},
  \bibinfo {author} {\bibfnamefont {L.}~\bibnamefont {Grando}}, \ and\ \bibinfo
  {author} {\bibfnamefont {S.}~\bibnamefont {Peruzzo}},\ }\href {\doibase
  https://doi.org/10.1016/j.fusengdes.2017.03.098} {\bibfield  {journal}
  {\bibinfo  {journal} {Fusion Engineering and Design}\ }\textbf {\bibinfo
  {volume} {123}},\ \bibinfo {pages} {892 } (\bibinfo {year}
  {2017}{\natexlab{b}})},\ \bibinfo {note} {proceedings of the 29th Symposium
  on Fusion Technology (SOFT-29) Prague, Czech Republic, September 5-9,
  2016}\BibitemShut {NoStop}%
\bibitem [{axi(2016)}]{axi4_stream_fifo_2016}%
  \BibitemOpen
  \href {\small https://www.xilinx.com/../pg080-axi-fifo-mm-s.pdf} {\enquote
  {\bibinfo {title} {Xilinx pg080 - axi4-stream fifo v4.1},}\ } (\bibinfo
  {year} {2016})\BibitemShut {NoStop}%
\end{thebibliography}%


\end{document}